\documentclass[10pt,twocolumn,letterpaper]{article}

\usepackage{iccv}
\usepackage{times}
\usepackage{epsfig}
\usepackage{graphicx}
\usepackage{amsmath}
\usepackage{amssymb}

\usepackage[breaklinks=true,bookmarks=false]{hyperref}

\iccvfinalcopy % *** Uncomment this line for the final submission
\def\iccvPaperID{****} % *** Enter the ICCV Paper ID here

\ificcvfinal\fi

\begin{document}

%%%%%%%%% TITLE
\title{RRScell method for automated single-cell profiling of multiplexed immunofluorescence cancer tissue}

\author{
  Alvason~Zhenhua~Li,
  Karsten~Eichholz,
  Anton~Sholukh,\\
  Daniel~Stone,
  Michelle~Loprieno,
  Keith~Jerome,
  %Jin\&Sun,\\
  Khamsone Phasouk,
  Kurt~Diem, Jia~Zhu,
  Lawrence~Corey\\
  \\
  Vaccine and Infectious Disease Division, Fred Hutchinson Cancer Research Center\\
Seattle, WA 98109, USA\\
{\tt\small alvali@fredhutch.org}
% For a paper whose authors are all at the same institution,
% omit the following lines up until the closing ``}''.
% Additional authors and addresses can be added with ``\and'',
% just like the second author.
% To save space, use either the email address or home page, not both
}

\maketitle
% Remove page # from the first page of camera-ready.
\ificcvfinal\thispagestyle{empty}\fi

%%%%%%%%% ABSTRACT
\begin{abstract}

Multiplexed immuno-fluorescence tissue imaging, allowing simultaneous detection of molecular properties of cells, is an essential tool for characterizing the complex cellular mechanisms in translational research and clinical practice.
  New image analysis approaches are needed because tissue section stained with a mixture of protein, DNA and RNA biomarkers are introducing various complexities, including spurious edges due to fluorescent staining artifacts between touching or overlapping cells.
  We have developed the RRScell method harnessing the stochastic random-reaction-seed (RRS) algorithm and deep neural learning U-net to extract single-cell resolution profiling-map of gene expression over a million cells tissue section accurately and automatically. 
  Furthermore, with the use of manifold learning technique UMAP for cell phenotype cluster analysis, the AI-driven RRScell has equipped with a marker-based image cytometry analysis tool (markerUMAP) in quantifying spatial distribution of cell phenotypes from tissue images with a mixture of biomarkers.
  The results achieved in this study suggest that RRScell provides a robust enough way for extracting cytometric single cell morphology as well as biomarker content in various tissue types, while the build-in markerUMAP tool secures the efficiency of dimension reduction, making it viable as a general tool in the spatial analysis of high dimensional tissue image.

\end{abstract}

%%%%%%%%% BODY TEXT
\section{Introduction}

Immuno-fluorescent staining is widely used in biomedical research to visualize particular cellular events by specific molecular markers, such as protein-based immuno-histochemistry (IHC), and DNA/RNA-based in situ hybridization (ISH). Recent advances in multiplexed IHC/ISH tissue imaging techniques, allowing simultaneous detection of numerous markers, raise the possibility of conducting deeper immune response profiling at a single-cell level \cite{Tan2020View, Eichholz2020carRNA}.

This growing field reveals a wealth of information about the relationships among cells in the entire tissue micro-environment but presents a significant image analysis challenges. For instance, the membrane-image quality defection due to the immuno-staining techniques is common and significant. Noises bring obstacles to edge detection because it reduces the contrast of real membrane boundary and also introduce spurious edges due to noisy contrast.
%as shown in Figure~{\ref{figure_introduction}}. 
%%%
% \begin{figure*}
%   \centering
%     \includegraphics[width = 0.5\linewidth]{figure/figure_introduction}
%     \caption{Challenges in analysis of multiplexed immuno-fluorescent images of cancer tissue with wide variance of intensity and noise.} 
%       (A) is a typical merged immuno-fluorescent stitched image of a whole cancer tissue slide (red color channel is RNA straining, green color channel is membrane straining, and blue color channel is nuclei staining).
%       (B1) is the raw image from membrane straining channel only.
%       (B2) is the raw image from nuclei staining channel only. Here, the messy zone is missing the nuclei straining.
%   \label{figure_introduction}
% \end{figure*}
%%%
Furthermore, complex tissue involved in diverse physiological processes, no single technique can provide all of the answers, so it’s necessary to use a combination of IHC and ISH methods to providing insights into physiological processes and disease pathogenesis, wherein a mixture of protein, DNA and RNA markers is introducing various complexities. Thus, the multiplexed image-based transcriptional profiling of cells is significant challenging. The cutting edge machine-learning approaches are the promising way to address the challenges. 

\subsection{Problems and Contributions}
The major drawback of supervised machine-learning techniques is that the quality of the results depends upon the amounts and quality of manually annotated training data.
However, the performance of human annotators show a high degrees of variability in complex fluorescent image in tissue as shown in Figure~{\ref{figure_ground_truth}(C)}. Furthermore, there is existing gold standard paradox in biomedical tissue \cite{Aeffner2017GoldStandard}, so that, it is not practical using manual annotation to generate the consensus ground truth in complex tissue image.

In this work, we aim to solve the practical problem of lacking ground-truth in microscopy tissue image, we have harnessed the stochastic random-reaction-seed (RRS) algorithm as a ground-truth generator. Then, we build a AI-driven pipeline U-net to extract single-cell level information from complex tissue.
Finally, we address the practical problem of large amounts of single-cell data output from the machine-learning pipeline,  demonstrate the simple cost-efficient approach to leverage the power of UMAP (Uniform Manifold Approximation and Projection) \cite{Mcinnes2018Umap}.

Specifically, the main contributions are as the following:
(1) We have designed a RRS-based automated ground-truth generator for training the neural networks, and importantly, the fine-tuned parameters of RRS ensured the generation of a high-quality ground-truth for microscopy image segmentation.
(2) We have developed an automated RRScell pipeline for extracting single-cell level profiling map of whole slide tissue. It is a robust convolutional network in biomedical image segmentation with limited training dataset.
(3) RRScell has a build-in markerUMAP tool secures the efficiency of dimension reduction, making it viable as a general tool in the spatial analysis of high dimensional tissue image. 

\subsection{Related Works}
Recent advances in machine-learning have outperformed the state of the art of traditional image analysis in many applications.
Supervised learning segmentation approaches require abundant ground-truth labels to deliver accurate inference. However, in biomedical microscopy image segmentation, these labor-intensive annotation are scarce \cite{Ke2020LazyLabel}. 
Among various deep convolutional encoder-decoder networks for image segmentation, the U-net is considered a generic architecture in biomedical cell segmentation \cite{Ronneberger2015Unet, Falk2019Unet} because data-augmentation strategy in U-net is able to achieve desired accuracy with
a few of ground-truth annotation.
There has been considerable efforts in computational multiplexed tissue image analysis,
some are commercial microscopy platforms or licensed software \cite{Czech2019cytoKit, Tan2020View}, so they are not easy to be understood, and customized for complex tissues. 
some are implemented in Matlab or Python wrapper for Matlab, and applied probability maps of nuclei/membrane for generating segmentation ground-truth masks \cite{Schapiro2017histoCAT, Czech2019cytoKit, Stoltzfus2020cytoMAP}.
Many existing multiplexed analysis pipelines are developed for specific assays or certain IHC/ISH schemes, there are lacking consensus standards for individual labs' bespoke IHC/ISH scheme. 
Image-based cytometry involves a wide range of cutting-edge techniques, more efforts are needed to overcome various issues and limitations in this active research field.

\section{Methods}
\subsection{Automated ground truth generator}
\begin{figure*}
  \centering
    \includegraphics[width = 0.6\textwidth]{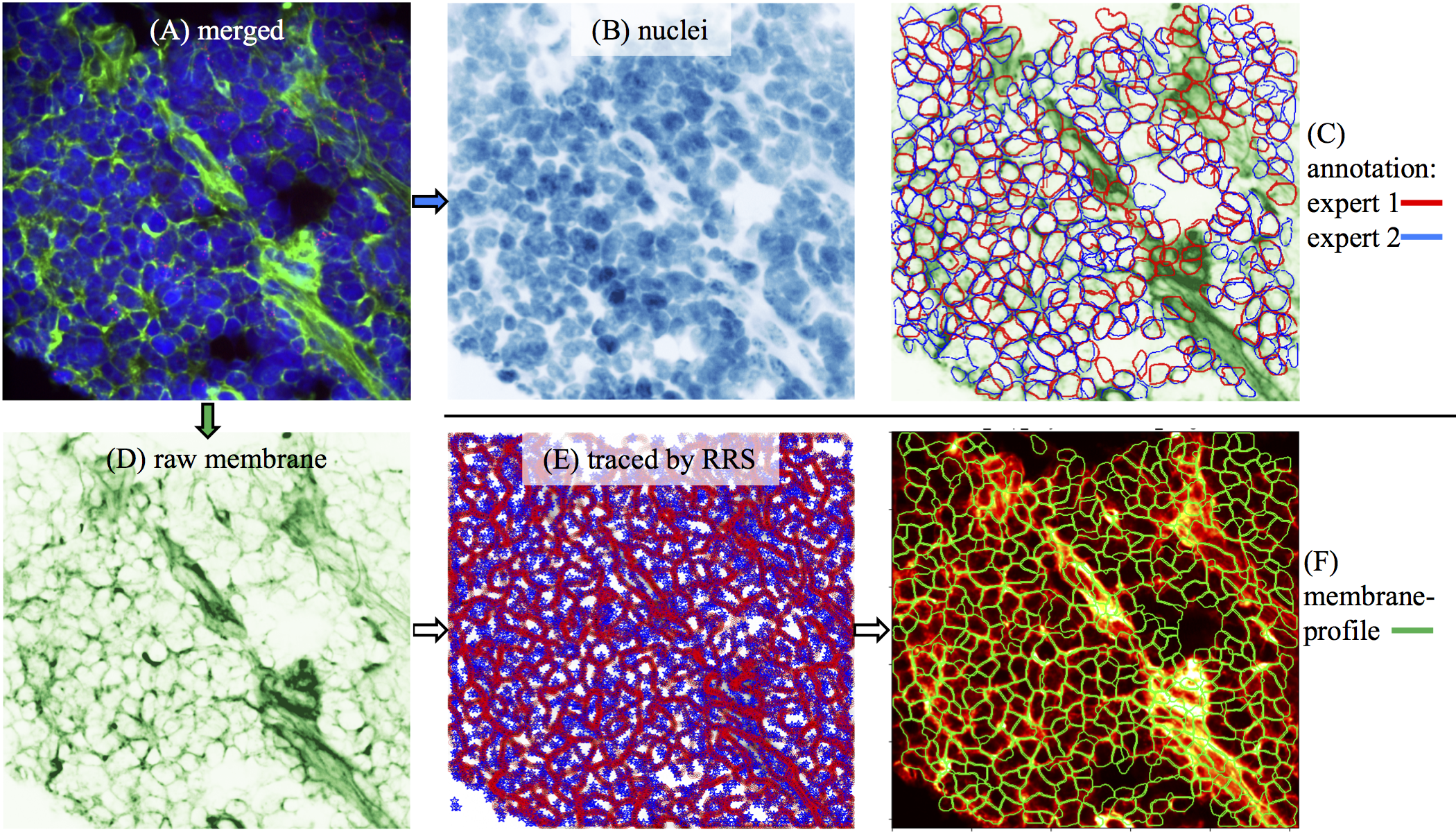}
    \caption{Automated ground truth generation in a sample image from cancer tissue.
      (A) is the merged image of multiplexed.
      (B) is raw nuclei channel.
      (C) are the membrane annotations from two experts (red color from one expert, blue color from another expert) on the same membrane image.
      (D) is the raw nuclei channel.
      (E) is the dynamic tracing map on the membrane image by the Random-Reaction-Seed method, where the random initial seed is denoted by blue star, the search chain is denoted by red-circle. \cite{Li2019RRS}.
      (F) is the traced membrane profile by RRS, the green-color is the traced membrane plotted on the raw membrane image (hot-color-map).
}
  \label{figure_ground_truth}
\end{figure*}

We have applied the stochastic RRS method \cite{Li2019RRS} to generate high quality ground-truth automatically, as shown in Figure~{\ref{figure_ground_truth}(E)}. The accuracy of automated ground-truth from RRS is ensured by handcraft parameters for selected image dataset.
%%%
%%%  
There is no objective  ground truth available for multiplexed immunofluorescence images. To assess the accuracy of our method we compare the results with the manual measurement.
%%%
\begin{figure*}
  \centering
 \includegraphics[width = 0.5\textwidth]{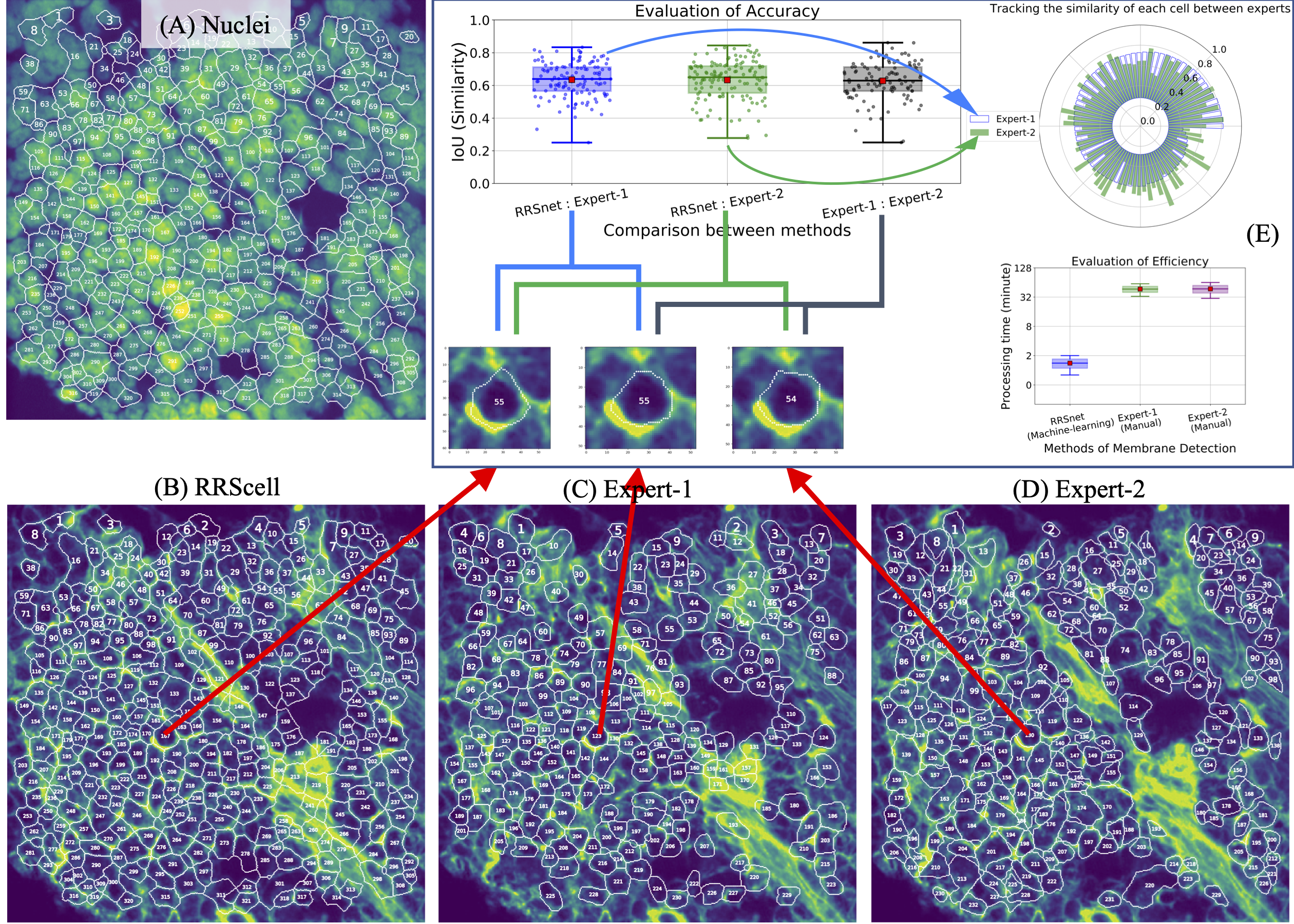}
 \caption{Performance of RRScell method.
   (A) Membrane profile from RRScell is plotted on the nuclei image.
   (B) Membrane profile from RRScell is plotted on the membrane image.
   (C) Manual membrane profile from Expert-1 is plotted on the membrane image.
   (D) Manual membrane profile from Expert-2 is plotted on the membrane image.
   (E) The box-plot on the top is the evaluation of accuracy among three different methods, the red-square-point inside each box is the mean value.
   Comparison between membrane-detection methods:
   the blue-color box is the similarity between RRScell and expert-1,
   the blue-color box is the similarity between RRScell and expert-2,
   the black-color box is the similarity between two experts.
   The polar plot of bar-chart on the upper-right is the evaluation of similarity of each cell between experts: each blue-color bar is corresponding to each dot in the blue-color box-plot, here, the bars are sorted in descending order.
   The box-plot on the lower-right is the evaluation of efficiency, the efficiency of RRScell is significantly better than the manual process (>30 folds difference in processing time).
 }
  \label{figure_performance}
\end{figure*}
%%%
As demonstrated in Figure~{\ref{figure_performance}}(E), the accuracy of RRScell method is equivalent to manual annotation. Furthermore, due to DAPI straining defect in high intensity zone beyond human visual approach, so that both experts are avoiding these messy areas, as shown in Figure~{\ref{figure_performance}}(C, D), one significant advantages of RRScell is able to tracing more cells based on its powerful combination of both membrane and nuclei as detection, as shown in Figure~{\ref{figure_performance}}(B).
The correct detection of cellular membrane is verified by the nuclei plot in Figure~{\ref{figure_performance}}(A).
As shown in Figure~{\ref{figure_performance}}(E), the polar plot of bar-chart, where the polar axis is the similarity score, the blue-color bars are corresponding to the cells (blue dots) in the blue box-plot while the green-color bars are corresponding to the cells (green dots) in the green box-plot. In order to demonstrate the variability of manual annotations, the blue-color bars are sorted in descending order, and then the green-color bar is placed by tracking the similar cellular physical location. So that the similarity of each cell annotated by two experts are displayed in a contrast way. The similarity score of each cell membrane profile is calculated by the IoU (intersection over union).
Automated RRS-based ground truth generator is accurate and consistent. 
\subsection{RRScell pipeline}
\begin{figure*}
  \centering
 \includegraphics[width = 0.5\textwidth]{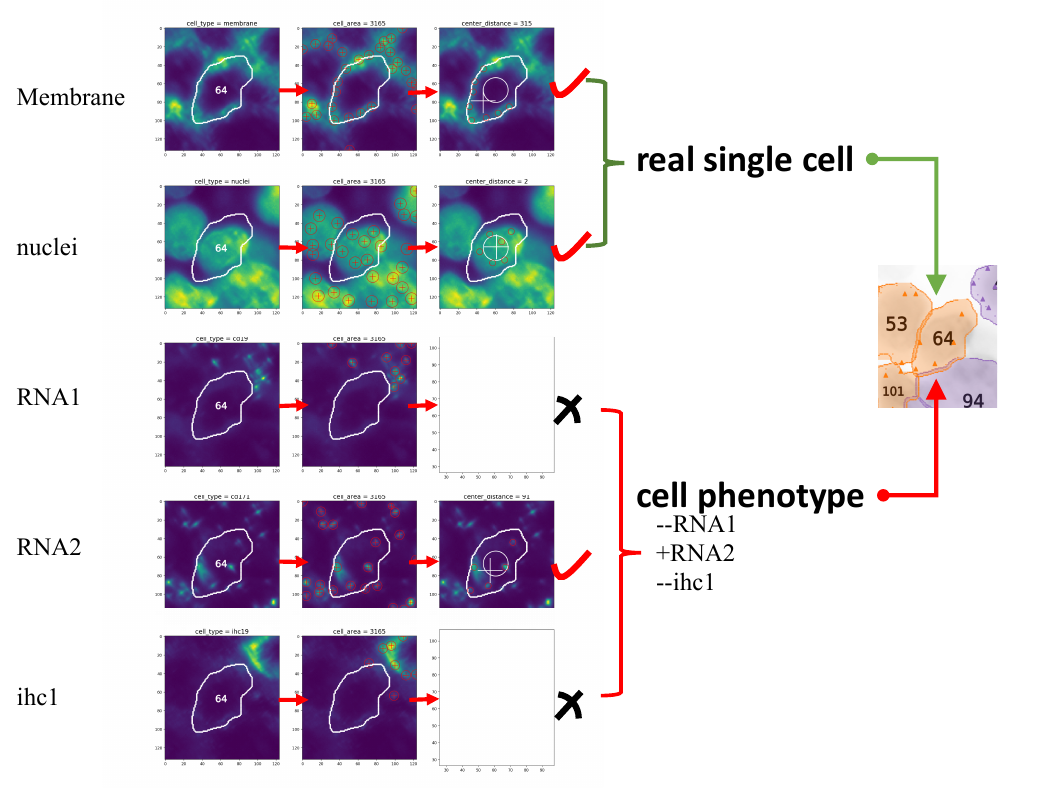}
  \caption{Scheme of automated cell phenotype detection. Here, an example of 5-plex image is presented: the initial step is the validation of real single cell based on both membrane and nuclei features, and a phenotype detection step will be followed in staining conditions (such as RNA1, RNA2, and IHC1).
}
  \label{figure_type_detection}
\end{figure*}

We have built an artificial intelligence (AI)-driven pipeline which is composed of the following modules:
(1) Automated cell segmentation is a U-net-based encoder-decoder network with RRS-based ground-truth generator. The cell segmentation is either based on raw membrane staining or anchored to synthetic/artificial membrane from raw nuclei staining/other suitable staining. 
(2) Automated cell-type detection: As shown in Figure~{\ref{figure_type_detection}}, we have developed a workflow for phenotype validation cell-by-cell.

\subsubsection{markerUMAP}
We have developed a marker-based image cytometry analysis tool (markerUMAP) in quantifying
spatial distribution of cell phenotypes from tissue images with a mixture of biomarkers.
The strategy of markerUMAP is introducing simplified symbolic marker to represent the captured biomarker contents. For example, the quantification of RNAscope measurement is focused on the detective RNA dot-like distribution instead of normal intensity measurement (since RNA staining enables specific and sensitive detection of RNA, which can be visualized as a dot, with each dot denotes a single RNA transcript). 
This symbolic marker-based strategy is organically compatible and consistent with UMAP fast, robust and meaningful organization of cell clusters \cite{Becht2019usingUmap}.

%%%
%%%

\section{Experiments and Results}
In this section, we applied RRScell into multiplexed tissue stained by various IHC/ISH schemes: starting from 3-plex to higher plexed immuno-fluorescence tissues in a gradual way.
\subsection{Three-plex CRISPR-Cas9 tissue images without membrane staining}
 In some practical translational research, multiplexed immunofluorescence tissue images are lacking membrane straining in many circumstances. Therefore, the RRScell pipeline is also equipped with the synthetic/artificial membrane generation step to work flexibly and consistenly on
 membrane-less images.
 As shown in Figure~{\ref{figure_RRScell_no_nuclei}, this immuno-fluorescent tissue image is from HBV-specific SaCas9 therapy \cite{Stone2021Cas9}.
 %%%
\begin{figure*}
  \centering
 \includegraphics[width = 0.6\textwidth]{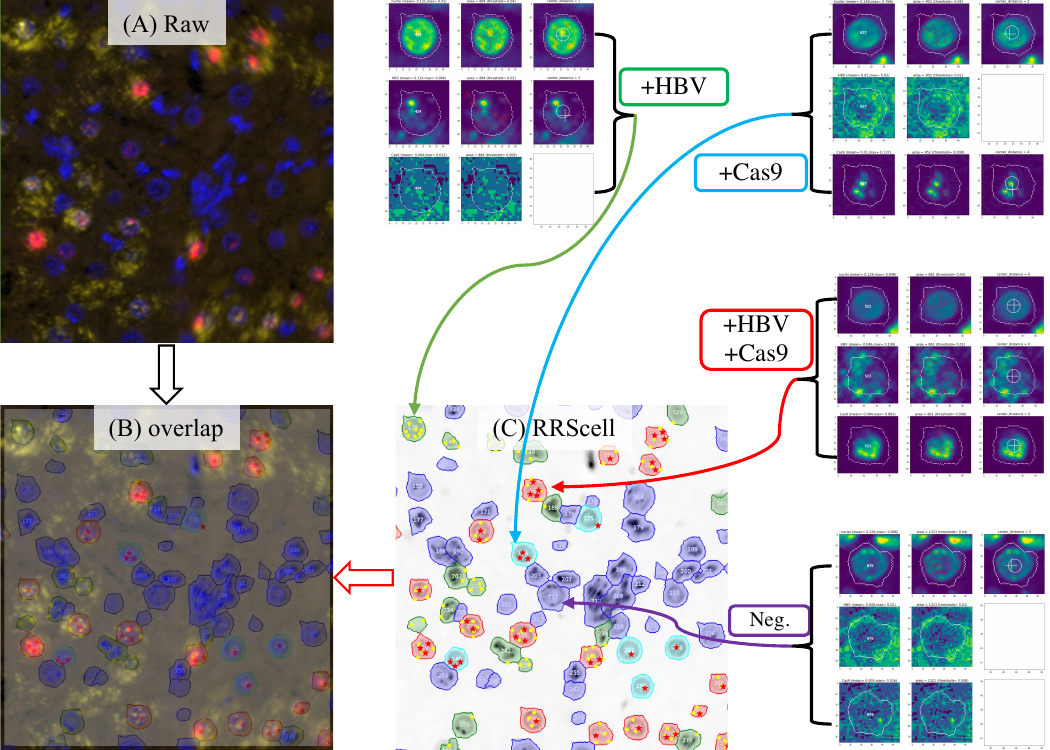}
 \caption{RRScell phenotype detection of 3-plex immuno-fluorescent tissue image from HBV-specific SaCas9 therapy. (A) is the raw image of 3 channels: red color denotes the Cas9 RNA-staining, yellow color denotes the HBV RNA-staining, and blue color denotes the nuclei staining. (B) is given the overlapped image from the raw and the RRScell profiling map. (C) is profiling map from the cell phenotype analysis by the RRScell pipeline: detected +HBV dots are denoted by yellow-disk, detected +Cas9 dots are denoted by red-star. There are four types of cells: +HBV, +Cas9, +HBV+Cas9, and negative -HBV-Cas9.
} 
  \label{figure_RRScell_no_nuclei}
\end{figure*}
%%%
%%%
\begin{figure*}
  \centering
 \includegraphics[width = 0.6\textwidth]{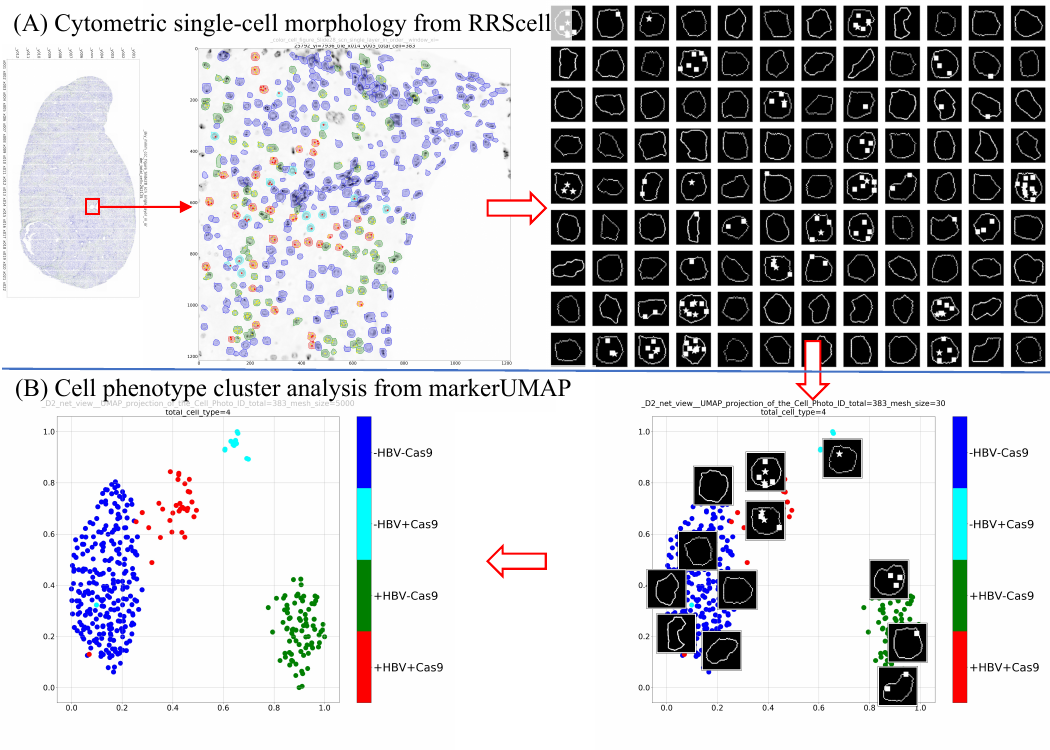}
  \caption{makerUMAP phenotype clustering analysis of 3-plex immuno-fluorescent tissue image from HBV-specific SaCas9 therapy. (A) Demonstration of extracting single-cell level information from individual cells in whole slide tissue by RRScell. (B) Demonstration of cell phenotype cluster analysis by markerUMAP tool. Note: only a small part of the tissue was applied in this demonstration.} 
  \label{figure_umap_3plex}
\end{figure*}
%%%
\subsection{Four-plex CAR-T RNA-scope tissue images with membrane staining}
Chimeric antigen receptor (CAR) T cells, as one of rapid emerging form of genetically engineered “artificial immune cell”, have outstanding therapeutic potential for treating cancers \cite{Eichholz2020carRNA}.
To study the anti-HIV CAR T cell trafficking into tissues with viral replication by microscopy, we developed an RNAscope- based RNA fluorescent in situ hybridization assay. In order to validate the RNA probes, we need to quantify a large volume of image dataset from in vitro culture and tissues sections among mouse, monkey, and human. However, the membrane-image quality defection due to the sugar-staining techniques is common and significant. Noises bring obstacles to edge detection because it reduces the contrast of real membrane boundary and also introduce spurious edges due to noisy contrast, as shown in Figure~{\ref{figure_result_4plex}.
%%%
\begin{figure*}
  \centering
 \includegraphics[width = 0.6\textwidth]{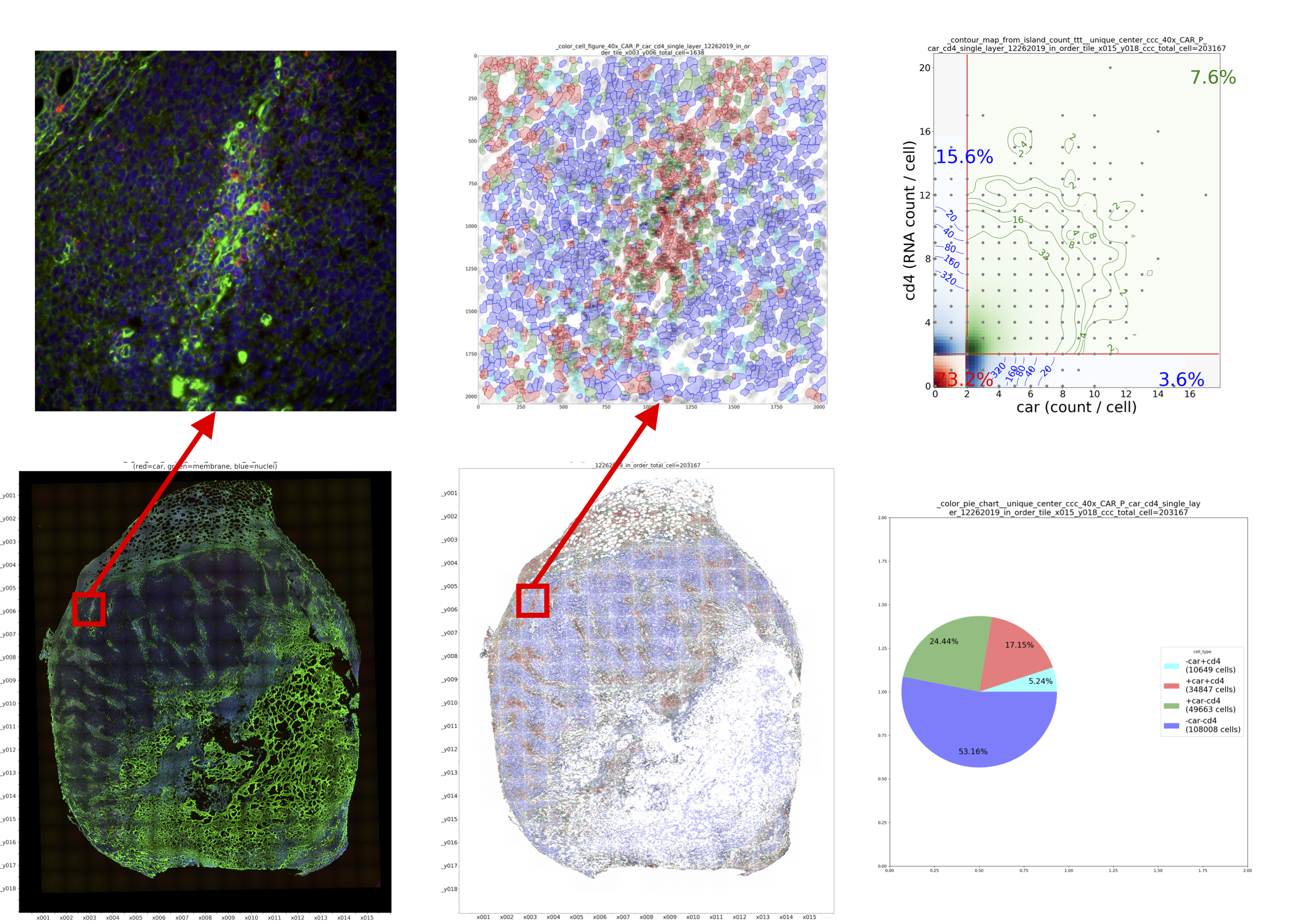}
  \caption{Cell phenotype analysis of 4-plex immuno-fluorescent tissue image stained by 4 biomarkers: CAR, CD4, nuclei,and membrane. The left panel is the raw stitched cancer tissue. The central panel is profiling map of 4 phenotypes from RRScell inferencing. The right panel is corresponding pie chart and the contour map with the gating condition of captured mRNA contents of individual cells from the whole slide tissue.
}
  \label{figure_result_4plex}
\end{figure*}
%%%

\subsection{Seven-plex cancer tissue images without membrane staining}
We are applied ChipCytometry platform for making highly multiplexed assay and imaging on cancer tissues. For higher multiplexed tissue, the potential total cell-phenotypes is in a exponential-growth pattern. For instance, there are total 34 cell-phenotypes are classified in a small corner of 7-plex tissue, as shown in Figure~{\ref{figure_umap_7plex}.
Significantly, the markerUMAP is able to provide the fast run times, high reproducibility and the meaningful organization of cell-pheonotype clusters, as shown in Figure~{\ref{figure_umap_cluster_on_tissue}.

%%%
\begin{figure*}
  \centering
 \includegraphics[width = 0.6\textwidth]{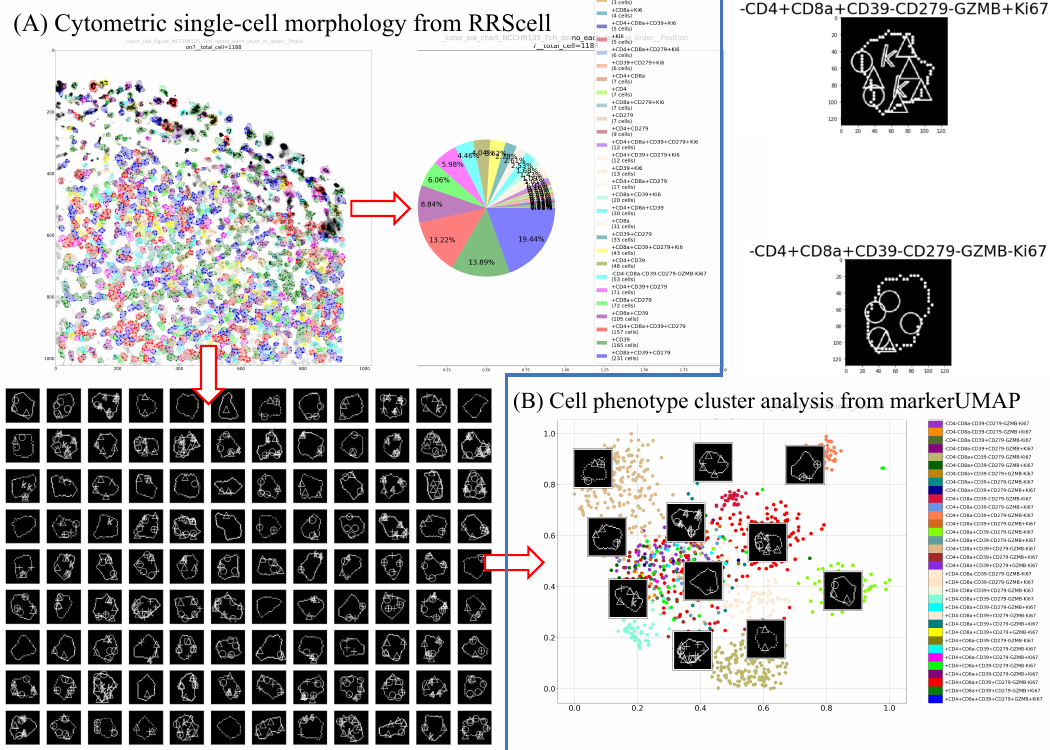}
 \caption{RRScell phenotype detection of 7-plex immuno-fluorescent image from cancer tissue stained by 7 biomarkers: CD4, CD8a, CD39, CD279, DAPI, GZMB, and Ki67. (A) Demonstration of extracting single-cell level information from individual cells in one corner of whole slide of tissue by RRScell. The pie-chart illustrates that a total of 34 cell-phenotypes are detected in the demo image. (B) Demonstration of cell phenotype cluster analysis by markerUMAP tool.  
} 
 \label{figure_umap_7plex}
\end{figure*}
%%%

%%%
\begin{figure*}
  \centering
 \includegraphics[width = 0.6\textwidth]{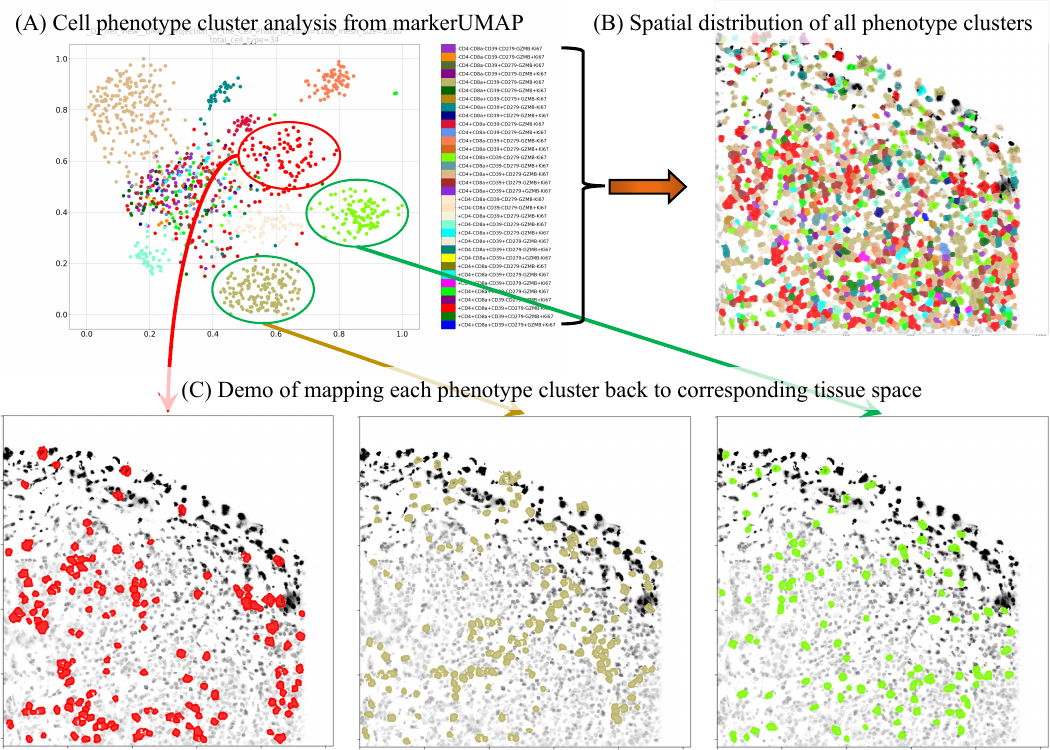}
 \caption{Demo of spatial analysis of high dimensional multiplexed images from cancer tissue stained by 7 biomarkers: CD4, CD8a, CD39, CD279, DAPI, GZMB, and Ki67. (A) is the cell phenotype cluster analysis by markerUMAP. (B) is spatial distribution of all phenotype clusters. (C) are demonstrations of mapping each phenotype cluster back to corresponding tissue space.
} 
  \label{figure_umap_cluster_on_tissue}
\end{figure*}
%%%

\section{Conclusion}
We present a robust, open-source Python pipeline for immune cell profiling in spurious-edge-tissues of translational medical research. This machine-learning solution builds upon RRS automated ground truth generation. The house-made automated machine learning pipeline is automatic scaling up from 3 to 7, or highly multiplexed immuno-fluorescent tissue images.
The AI-driven RRScell method has demonstrated the robustness and flexibility for single-cell profiling of various tissue types and formats where membrane and nuclei are often missed.

In conclusion, RRScell, as a framework combining HMM-based (hidden-Markov-model) RRS, deep neural learning network U-net, and manifold learning technique UMAP, has demonstrated the effectiveness and robustness for precise quantification of immune cell phenotypes in images taken from noisy multiplexed immunofluorescence cancer tissue, which is becoming an increasingly useful tool for biomedical researches owing to its ability to profiling cellular phenotyping in cancer resection tissues with various complexities. Furthermore, the markerUMAP image-based cytometry tool inside RRScell has emerged as promising spatial analysis tool in ongoing high-dimensional immuno-straning tissues in pre-clinical research and clinical trials.

%%% use section* for acknowledgment
% \section*{Acknowledgments}
% This work was partially supported by the National Institutes of Health [grant number UMI A126623].
% The authors would like to thank Drs. Sijie~Sun, and Mindy~Miner for their constructive comments to improve the quality of the paper.
% All the raw fluorescent images are provided by Karsten~Eichholz in the HSV Vaccine Lab at the Vaccine and Infectious Disease Division of Fred Hutchinson Cancer Research Center.

% \section*{Author contributions statement}
% A.Z.L. conceived the algorithm, analyzed the results and wrote the manuscript; E.K. coordinated and developed the chimeric antigen receptor T cells investigations; A.S. coordinated the setup of GPU client-server; L.C. supervised the cancer tissue studies.

% \section*{Data availability statement}
% The datasets generated and analyzed during the current study are available from the corresponding author on reasonable request.

% \section*{Additional information}
% \textbf{Competing interests:} The authors declare no competing interests.

%%% ---

% -------------------------------------------------------------------------
%%%
{\small
\bibliographystyle{ieee_fullname}
\bibliography{RRS_unet_umap_cell_Reference}
}

\end{document}